# Microcavity induced by few-layer GaSe crystal on silicon photonic crystal waveguide for efficient optical frequency conversion

Xiaoqing Chen[1], Yanyan Zhang[2], Yingke Ji[1], Yu Zhang[1], Jianguo Wang[1], Xianghu Wu[1], Chenyang Zhao[3], Liang Fang[1], Biqiang Jiang[1], Jianlin Zhao[1] and Xuetao Gan[1,4,*]

[1]Key Laboratory of Light Field Manipulation and Information Acquisition, Ministry of Industry and Information Technology, and Shaanxi Key Laboratory of Optical Information Technology, School of Physical Science and Technology, Northwestern Polytechnical University, Xi'an 710129, China
[2]School of Artificial Intelligence, OPtics and ElectroNics (iOPEN), Northwestern Polytechnical University, Xi'an 710072, China
[3]Analytical & Testing Center, Northwestern Polytechnical University, Shaanxi 710072, China
[4]School of Microelectronics, Northwestern Polytechnical University, Xi'an 710129, China

**Abstract**: We demonstrate the post-induction of high-quality microcavity on silicon photonic crystal (PC) waveguide by integrating few-layer GaSe crystal, which promises highly efficient on-chip optical frequency conversions. The integration of GaSe shifts the dispersion bands of the PC waveguide mode into the bandgap, resulting in localized modes confined by the bare PC waveguides. Thanks to the small contrast of refractive index at the boundaries of microcavity, it is reliably to obtain quality (Q) factors exceeding $10^4$. With the enhanced light-GaSe interaction by the microcavity modes and GaSe's high second-order nonlinearity, remarkable second-harmonic generation (SHG) and sum-frequency generation (SFG) are achieved. A record-high on-chip SHG conversion efficiency of 131100% $W^{-1}$ is obtained, enabling the clear SHG imaging of the resonant modes with the pump of sub-milliwatts continuous-wave (CW) laser. Driven by a pump of on-resonance CW laser, strong SFGs are successfully carried out with the other pump of a CW laser spanning over the broad telecom-band. Broadband frequency conversion of an incoherent superluminescent light-emitting diode with low spectral power density is also realized in the integrated GaSe-PC waveguide. Our results are expected to provide new strategies for high-efficiency light-matter interactions, nonlinear photonics and light source generation in silicon photonic integrated circuits.

**Keywords**: Optical microcavity, photonic crystal, gallium selenide, second harmonic generation, sum frequency generation

## Introduction

Silicon photonic integrated circuits (PICs) are attractive to strengthen the developments of optical telecom, datacom, computing and sensing, benefiting from their CMOS-compatible manufacturing process in high-volume production at low cost.[1–3] Light source is one of the essential active components on PICs.[4] Limited by silicon's indirect bandgap, it is difficult to directly realize light emission on silicon PICs via carrier recombination. Optical frequency conversion (OFC) is another mechanism to obtain light emission relying on materials' intrinsic nonlinear optical responses.[5–7] Based on silicon's third-order nonlinearity, bright green light emitted by third-harmonic generation (THG) has been demonstrated from silicon photonic crystal (PC) waveguide.[5] This green light source has been employed to carry out a compact on-chip autocorrelator to measure the pulse width of a pulsed laser.[8] However, because of the extremely small third-order nonlinear susceptibility, moderately strong THG has to be pumped by pulsed lasers, even with the enhancement of slow-light effect.[5] To improve OFC on silicon photonics, microcavities have been exploited to enhance the light-matter interaction.[9–14] Assisted by a PC microcavity, which is one of the optical resonators with the highest ratio of the quality (Q) factor to the mode volume, THG pumped by continuous-wave (CW) laser was realized.[12] In our previous work, we integrated few-layer GaSe crystal on silicon PC microcavities, whose two-dimensional geometry is compatible with the planar slab of silicon photonics.[10] Different from silicon, GaSe has second-order nonlinearity due to the noncentrosymmetric crystal structure.[15] The value of second-order nonlinear susceptibility $\chi^{(2)}$ is about ten orders of magnitude higher than that of the third-order nonlinear susceptibility $\chi^{(3)}$.[16] Hence, the second-order nonlinear process is much stronger than the third-order nonlinear process when they are driven by the optical field within the strength range provided by a CW laser. Consequently, in the integrated GaSe-PC microcavity, we obtained light emission of second-harmonic generation (SHG) with a strength more than two orders of magnitude higher than that of the THG.[10] This integration architecture also supports the emission of interesting cascaded sum-frequency generation (SFG).[13] However, in these works, the integration of few-layer GaSe crystal degrades the Q factor of the PC microcavity because the high refractive index of the GaSe flake causes an asymmetric dielectric perturbation across the vertical direction of the silicon PC slab. Also, the utilized PC microcavities were designed with defects of shifted air-holes or missed air-holes, which have low tolerance to the fabrication imperfections.[17,18]

In this paper, we demonstrate that the integration of a few-layer GaSe crystal on a silicon PC waveguide could directly induce microcavities with Q factors exceeding $10^4$ and assist efficient on-chip OFCs with the pump of CW lasers. The integration of GaSe flake lowers the photonic transmission bands of the PC waveguide with the perturbation on the mode refractive index. Microcavity modes are obtained relying on the mode gap effects and further confirmed by the numerical calculation of the band diagrams of the GaSe-integrated and bare PC waveguides.[19–22] Since the GaSe crystal is part of the induced microcavity, this architecture is good at improving light-GaSe interaction through these high Q resonant modes. With the pump of on-resonance CW lasers, efficient SHGs are realized with a record-high conversion efficiency of 131100% $W^{-1}$, which can be imaged directly by a low sensitive camera. The high Q resonant modes of the GaSe-induced microcavity also promise broadband SFG pumped by two CW lasers, even though one of the two pump lasers is off-resonance from the microcavity modes. The number of resonant modes and Q factors of the GaSe-induced microcavities can be easily tuned by the coverage area and thickness of the GaSe layer. These findings would promote the development of high-efficiency nonlinear photonics and light sources on silicon PICs.[23,24]

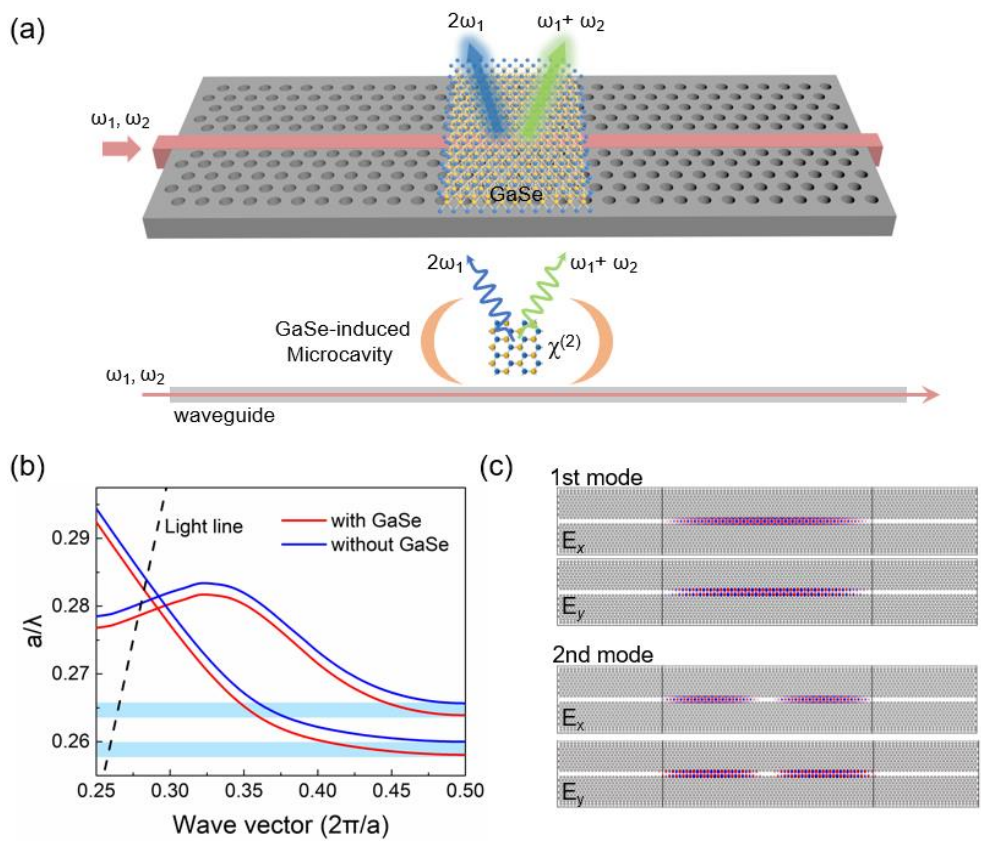

Fig. 1 | Schematic and theoretical calculations of the GaSe-induced microcavity on silicon PC waveguide. (a) Schematic of the GaSe-induced microcavity on silicon PC waveguide for efficient OFCs. Top panel: Few-layer GaSe crystal integrated on silicon PC waveguide. Bottom panel: Cavity induced by the GaSe flake and the cavity-enhanced OFCs

based on second-order nonlinearity of GaSe. (b) Calculated band structure of the line-defect PC waveguides integrated with (red solid line) and without (blue solid line) the few-layer GaSe crystal, where the dashed line represents the light-line of the silicon slab. The shadowed regions indicate the photonic bands supporting the emergence of resonant modes on GaSe-PC waveguide confined by the bare PC waveguide. (c) Mode distributions of the first and second order resonant modes of the GaSe-induced microcavities. The black solid lines indicate the area covered with the GaSe crystal.

## Results and discussion

Figure 1a illustrates the formation of GaSe-induced microcavity on the silicon PC waveguide and the OFCs enabled by the cavity enhanced light-mater interaction. The employed PC waveguide is formed by missing a line of air-holes in a triangle-lattice silicon PC slab. The few-layer GaSe crystal is integrated on the middle part of the PC waveguide, which separates the waveguide into three sequential sections, i.e., bare PC waveguide, GaSe-PC waveguide, bare PC waveguide. Because of the higher effective refractive index of the middle GaSe-PC waveguide compared with that of the bare PC waveguides at the two sides, light coupled into the section of GaSe-PC waveguide could be confined by the two side sections of bare PC waveguides, which gives rise to a microcavity. To illustrate that, we simulate the dispersion curves of the waveguiding modes of the GaSe-PC waveguide and bare PC waveguide. Here, the PC lattice constant is $a = 410$ nm and the air-hole radius is $r = 120$ nm. The refractive indices of silicon and GaSe are chosen as 3.47 and 2.5, respectively. The two nearest neighbor air-hole rows surrounding the PC waveguide are shifted away from the line defect by 60 nm. The PC waveguide has two types of propagation mode within the photonic bandgap, recognized as even mode and odd mode.[19,22] As shown in Figure 1b, the integration of GaSe moves the dispersion curves of the waveguiding modes to lower regions with respect to the bare PC waveguide. As a consequence, the guiding modes of the GaSe-PC waveguide locating in the shadow region of Figure 1b cannot propagate through the bare PC waveguide because they are out of the transmission bands of the bare PC waveguide. The two side sections of bare PC waveguide could be considered as mirrors for the waveguiding mode in the GaSe-PC waveguide, which provides the mode confinement along the waveguide. In the in-plane direction perpendicular to the waveguide, the lattice of periodic air-holes would confine guiding modes of the GaSe-PC waveguide since their dispersion band locates inside the photonic bandgap of the PC lattice. With these optical confinements, the GaSe-PC waveguide region could be considered as a microcavity. Calculated by the finite element method, the two lowest order resonant modes are found, which are consistent with the mode-gap difference region located below the odd transmission band of the bare PC waveguide. These resonant modes have the calculated $Q$ factors in the order of $10^5$. Figure 1c gives the distributions of the $x$ ($E_x$) and $y$ ($E_y$) components of the electric field in the corresponding mode. The results illustrate the resonant modes localized in the central GaSe-covered region are formed by the standing-wave condition with reflections from the bare PC waveguide sections on the two sides, and the resonant modes with different orders have varied wave packets.

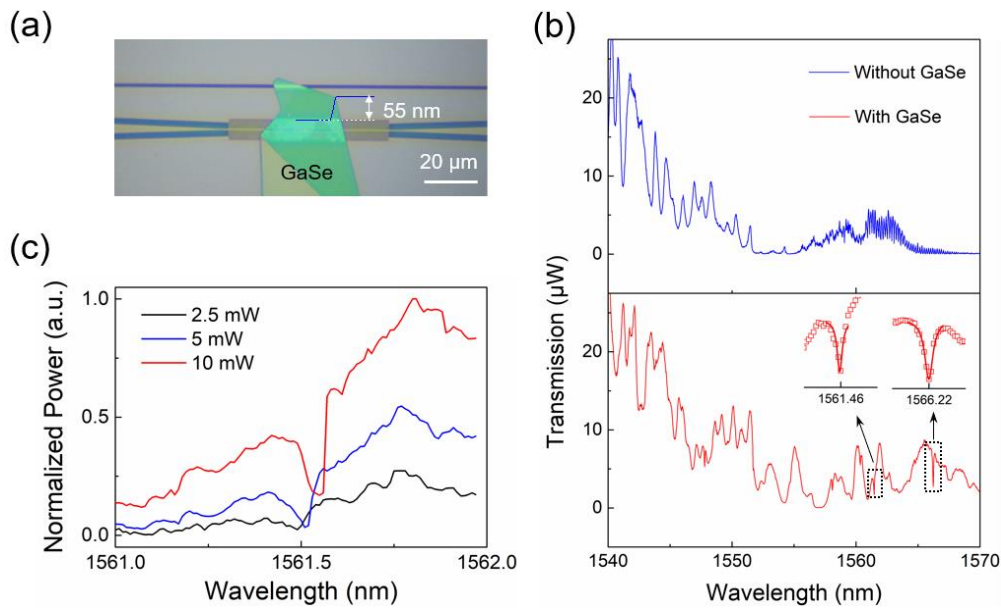

Fig. 2 | GaSe-induced microcavities in the silicon PC waveguide. (a) Optical microscope image of the fabricated device. (b) Transmission spectra of the PC waveguide before and after the integration of the GaSe crystal. Resonant modes of the induced microcavity are indicated in the zoomed insets. (c) Transmission spectra around the resonant mode of 1561.4 nm with the incident laser powers at 2.5 mW, 5 mW and 10 mW.

In the experiment, we fabricate the PC waveguides in a standard silicon-on-insulator wafer with a 220-nm-thick top silicon layer by electron-beam lithography and inductively coupled plasma etching. For the light coupling, the PC waveguide is connected with single-mode strip waveguides (with a width of 500 nm). Grating couplers are designed at the ends of the strip waveguides to facilitate the light coupling from optical fiber to the waveguide, which has a broad transmission range from 1500 nm to 1600 nm (Figure S1 in the Supporting Information).[25] More details of the device fabrication and characterization are presented in the Methods. The few-layer GaSe crystals mechanically exfoliated from a bulk crystal are transferred onto the central part of the silicon PC waveguide through the PDMS-assisted dry transfer method.[26,27] Figure 2a shows the optical microscope image of one of the fabricated devices. The thickness of this GaSe flake is 55 nm as characterized by the atom force microscope (AFM). The length of the PC waveguide is 61 μm with GaSe covering the central region of a length up to 29.5 μm. The facile Van der Waals integration of the two-dimensional GaSe crystal requires no precise alignment with the PC waveguide.[22,28]

The fabricated devices are characterized by measuring their transmission spectra. A tunable narrowband laser is employed as the light source to couple into the waveguides, and the transmitted powers at the waveguide output are monitored by a power meter. Figure 2b shows the transmission spectra in the range from 1540 nm to 1570 nm of the PC waveguide before and after the integration of the GaSe crystal. For the bare PC waveguide, clear Fabry-Perot (FP) oscillations are observed because the boundaries between the PC waveguide and strip waveguides form two mirrors. The oscillation periods become smaller towards the longer wavelength due to the slow light effect. After the integration of GaSe crystal, significant modifications occur due to the perturbation of the waveguide mode dispersions. The transmission spectrum shows a redshift about 4.5 nm due to the lowering of the transmission bands (see details in Figure S2 in the Supporting Information). We note that there are several dips in the transmission spectrum of GaSe-PC waveguide, for example locating at 1561.4 and 1566.2 nm, as zoomed in the insets of Figure 2b. These dips hint the existence of resonant modes coupled with the PC waveguiding mode,[20,22] which results from the microcavity induced by the integration of GaSe, as demonstrated in Figure 1. Here, the resonant mode at 1561.4 nm (1566.2 nm) is fitted by a Lorentzian lineshape, indicating a high $Q$ factor as 28900 (18400). Similar as the discussions in our previous work,[22] the induced resonant mode of the microcavity could be verified by the shift of the resonant wavelength with the increment of the input power. As shown in Figure 2c, the resonant dips at 1561.4 nm are acquired at varied input powers of 2.5 mW, 5mW and 10 mW. Because of the two-photon absorption in silicon at the telecom-band, the microcavity would be heated at high laser powers governed by the enhanced light-matter interactions in the resonant modes, giving rise to the red-shifted resonant wavelength due to the thermo-optic effect in the silicon and GaSe.

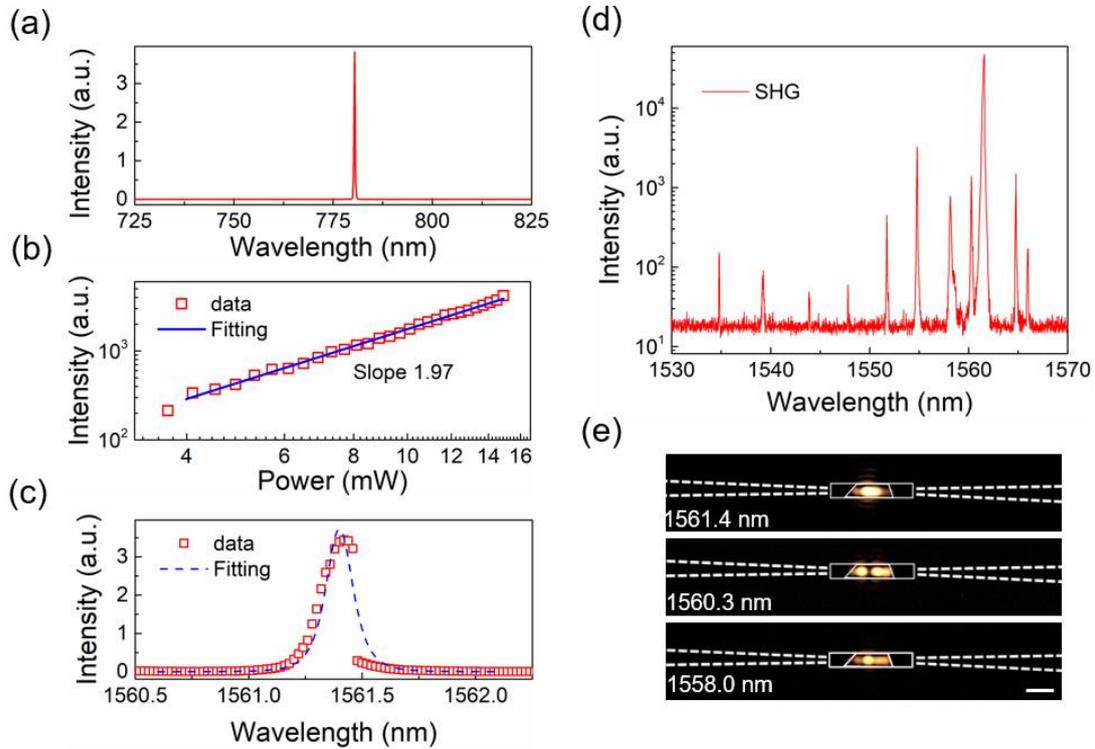

Fig. 3 | Enhanced SHG by the GaSe-induced microcavities. (a) SHG spectrum acquired with the pump at the resonant wavelength of 1561.4 nm (b) Log-log plot of the SHG powers versus pump powers fitted with a line in a slope of 1.97. (c) Measured SHG powers when the pump wavelength is scanned around 1561.4 nm, which can be fitted with a square of a Lorentzian lineshape. (d) Measured SHG powers when the pump wavelength is scanned over the range from 1530 to 1570 nm. (e) Images of SHGs from GaSe-PC waveguide pumped with CW lasers on resonant modes at 1561.4 nm, 1560.3 nm and 1558 nm. The scale bar is 20 μm.

As part of the induced microcavity, the GaSe flake couples with the localized resonant modes effectively. From the simulation results shown in Figure 1c, the mode distributed inside the GaSe crystal is calculated about 8.9% of the whole resonant mode. In addition, GaSe has the considerably strong second-order nonlinearity, which is about one order of magnitude higher than the commonly employed lithium niobate at the telecom-band.[29] With the enhanced light-GaSe interaction by the resonant modes with high density of photon state, high-efficiency OFCs based on GaSe's second-order nonlinearity are expected from the GaSe-integrated PC waveguide. To examine that, we couple a CW laser at the wavelength of 1561.4 nm to match with the transmission dip shown in Figure 2b. The light scattering from the GaSe region covered on PC waveguide is collected using an optical fiber from its top surface, which is then analyzed by a spectrometer mounted with a cooled silicon camera. The acquired spectrum in the near-infrared range is shown in Figure 3a. A strong signal at the wavelength of 780.7 nm is observed clearly. It corresponds to the half of the pump wavelength, indicating the occurrence of SHG. To confirm that, the power dependence of the signal at 780.7 nm on the pump power is measured, as shown in the log-log plot of Figure 3b. The data is fitted with a slope of 1.97. This quadratic function of the powers proves the SHG process that two photons of the fundamental pump wave are converted into one photon of the SHG wave.

To further verify the measured SHG signal is the result of the enhancement by the microcavity, we scan the laser wavelength around 1561.4 nm and monitor the top scattering SHG signal from the GaSe region. The measured result is shown in Figure 3c. The scattering SHG signal has the maximum intensity with the pump laser locating at 1561.4 nm. When the laser is tuned away from 1561.4 nm, the SHG signal reduces sharply to unobservable values. The wavelength dependence of the SHG signal is consistent with the lineshape determined by the resonant mode shown in Figure 2b. In the SHG process, the second-harmonic polarization $P(2\omega)$ is proportional to the square of the electric field $E(\omega)$ in the pump laser with the frequency of $\omega$. For the resonant mode at 1561.4 nm, its electric field $E(\omega)$ has a Lorentzian type lineshape, as described in the fitting in Figure 2b. Hence, the lineshape of the SHG signal on the pump wavelength shown in Figure 3c should be governed by the square of the Lorentzian lineshape in Figure 2b, which is calculated and displayed as the dashed line in Figure 3c. It shows good agreement with the experimentally measured result.

After the demonstration of the microcavity enhanced SHG in the GaSe-PC waveguide, we tune the wavelength of the pump laser from 1530 nm to 1570 nm to study the SHG process driven by other resonant modes. The measured scattering SHG signals are shown in Figure 3d. Interestingly, while only two resonant modes at 1561.4 nm and 1566.2 nm are observed clearly via the transmission dips in Figure 2b, there are several SHG peaks emerging when the pump laser is tuned to other wavelengths. In the linear transmission of the PC waveguide, the extinction ratio of transmission dips of the resonant modes coupled with the waveguiding mode is determined by the relationship between their coupling strength and the intra-cavity intrinsic loss.[30] Therefore, many resonant modes of the induced microcavity are not observed in Figure 2b. In contrast, the measured SHG signals are pumped by the resonant modes in the induced microcavity locally, which are scattered from the GaSe region directly and separated from the linear transmission mode in the PC waveguide for eliminating the destructive interference. Hence, the SHG signals are clear. We note that this result also implies a valid technique to distinguish resonant modes in micro-resonators, even for dark modes without far-field scattering or transmission.[31]

As demonstrated above that the pump wavelength dependence of SHG signal is a quadratic function of the lineshape of resonant mode,[23] the $Q$ factor of the emerged resonant modes could be extracted from the lineshapes of the SHG spectra. The $Q$ factors of these resonant modes indicated by the SHG peaks in Figure 3d are estimated in the scale of $10^4$. The experimentally measured values are lower than the theoretical predictions, which could be attributed to the fabrication imperfections of the silicon PC waveguide. As explained in Figure 1, the microcavity induced by the GaSe crystal is equivalent to a FP type resonator. The free spectral range ($FSR$) between the resonant modes is determined by $\lambda_0^2/(2n_{eff}L)$, where $\lambda_0$ is the resonance wavelength, $n_{eff}$ is the effective refractive index of the GaSe-PC waveguide mode, $L$ is the cavity length (length of the GaSe covered region). Around $\lambda_0 = 1554.7$ nm, the measured $FSR$ = 3.3 nm, and $L$ = 29.5 μm, determines the effective refractive index $n_{eff} \approx 12.4$. The small contrast of the effective refractive indices between the bare PC waveguide and GaSe-PC waveguide is favorable for suppressing radiation loss caused by the abrupt changes of guiding modes, which is key to achieve large $Q$ factors.[21]

SHG from the GaSe induced microcavity has very high efficiency, which is determined by the high $Q$ factor of the resonant mode, effective mode coupling with the GaSe crystal, and high second-order nonlinearity of GaSe. In Figure 3e, we display the SHG images captured from the integrated devices when it is pumped by on-resonance CW lasers. Here, an ordinary silicon CMOS camera with a low sensitive as that in an iPhone is employed. Bright and uniform SHG signals are observed over the GaSe region, marked by the dashed lines, which also implies the assistance of GaSe in realizing SHG from a silicon photonic chip. In previously reported works about OFCs in silicon photonic chips or OFCs in two-dimensional materials, SHG signals are normally pumped by pulsed lasers and characterized by cooled CCD camera with high sensitivity or sensitive photomultiplier tube. Here, the clear detection of the SHG signal pumped with a CW laser using a low sensitive camera indicates the high OFC efficiency in the GaSe-PC waveguide. The SHG images also exclude the possibilities of the point-defect cavities that enhance the light field at the localized positions. We evaluate the conversion efficiency of the enhanced SHG by calibrating the SHG signal from the GaSe flake with the emission from a commercial 780 nm laser diode (See Methods). With a 10 mW power of the laser at 1561.4 nm coupling from the input grating coupler, the transmission power collected from the output grating coupler is about 1.5 µW. It gives the coupling efficiency of the couplers and indicates the pump power accessing to the GaSe-covered PC waveguide is about 122 µW. With that, the emitted SHG signal from the GaSe region is calibrated as 20 µW. As a result, a record-high conversion efficiency of 131100% $W^{-1}$ is realized. [6,9-11,13,32]

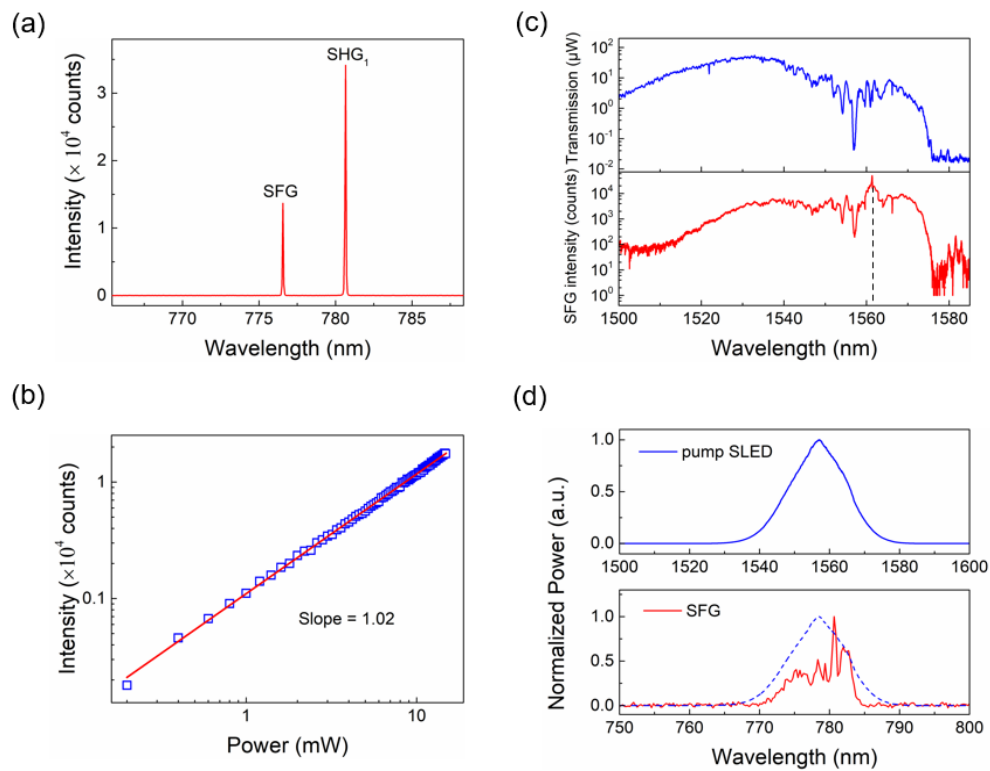

Fig. 4 | Broadband SFG from the GaSe-induced microcavity in the silicon PC waveguide pumped with an on-resonance CW laser and an off-resonance CW laser. (a) Spectrum of the SHG and SFG signals pumped by an on-resonance CW laser at 1561.4 nm and an off-resonance CW laser at 1545 nm. (b) Log-log plot of the SFG powers with the power of

Laser$_1$ at 1561.4 nm kept constant and the power of Laser$_2$ at 1545 nm ramping from 0.2 mW to 14.8 mW. The curve can be fitted linearly with a slope of 1.02. (c) Top panel: Transmission spectrum from 1500 nm to 1585 nm; Bottom panel: SFG powers with Laser$_1$ fixed at 1561.4 nm and Laser$_2$ scanned from 1500 nm to 1585 nm. The dashed line points out the position of Laser$_1$. (d) Top panel: Spectrum of the employed broadband SLED; Bottom panel: SFG spectrum pumped with the SLED. Halved spectrum of the employed broadband SLED is shown with the dashed line.

We then demonstrate the GaSe-induced microcavity on the silicon PC waveguide also supports efficient SFG, which is another important OFC process. Instead of two synchronous pulsed lasers utilized in other SFG architecture,[22,33–36] we employ two CW tunable narrowband lasers (Laser$_1$ and Laser$_2$) as pumps thanks to the strong cavity enhancement. When the two pump lasers are on-resonance with the microcavity modes, both are enhanced resonantly. Consequently, their SFG could have similar efficiency as that of the cavity-enhanced SHG process (See Figure S3 in the Supporting Information). As demonstrated in our previous work, the double-resonance pump could even support the interesting cascade SFG process.[13] Here, we would like to show that the enhancement on OFCs by the GaSe-induced microcavity is so strong that SFG could be achieved with the pump of one off-resonance CW laser. As shown in Figure 4a, by tuning the laser wavelengths of two CW pumps as 1561.4 nm (Laser$_1$) and 1545 nm (Laser$_2$), which are on-resonance and off-resonance respectively, two peaks at 780.7 nm and 776.5 nm are observed. According to the wavelength conversion, these peaks are recognized as SHG$_1$ of the on-resonance Laser$_1$ at 1561.4 nm and SFG of the two pump lasers. These two OFC processes are further confirmed by their pump power dependences. In the SFG process, it requires one photon from Laser$_1$ and one photon from Laser$_2$. Therefore, when only the power of Laser$_2$ varies and the power of Laser$_1$ is fixed, the SFG power should follow a linear function of the pump power, which agrees well with the fitting slope of 1.02 in Figure 4b. In the above OFCs, SHG (expected at 772.5 nm) pumped by the off-resonance Laser$_2$ at 1545 nm is not observed, which is attributed to its very low conversion efficiency without the cavity-enhancement. However, the SFG could be pumped efficiently by the off-resonance Laser$_2$ with the assistance of the on-resonance Laser$_1$. Moreover, the obtained intensity of SFG is comparable to that of the SHG pumped by the on-resonance $Laser_1$, illustrating that the field enhancement from the resonant mode of the GaSe-induced microcavity is extremely strong.

With the SFG character that only one of the pump lasers is required to be on-resonance, we further demonstrate that SFG of the GaSe-PC waveguide could be employed to upconvert light over the telecom-band into the near-infrared range. Similar as the operation in Figure 4a, the on-resonance $Laser_1$ at 1561.4 nm is fixed as one pump, and the other pump $Laser_2$ is the narrowband CW laser with the wavelength tunable over the 1500 nm to 1585 nm. The upconverted SFG signals versus the wavelengths of $Laser_2$ are recorded and shown in the bottom plot of Figure 4c. In the top plot of Figure 4c, the linear transmission spectrum of $Laser_2$ is also displayed for comparison. The wavelength dependence of the SFG signals have the same trend of the linear transmission of $Laser_2$, implying the efficient and broadband operation of the frequency upconversion process in the GaSe-integrated silicon PC waveguide with the assistance of another fixed on-resonance laser. This performance has potential applications in the octave-spanning bandwidth of on-chip optical signal processing.[37] The efficient OFC driven by one on-resonance pump laser also promises other expected processes, for instance difference frequency generation (DFG) and the resulted optical parametric amplifier. However, limited by our experimental setup and the small electronic bandgap (1.1 eV) of silicon, DFG signal locates in the terahertz range, which cannot be detected. In future, by changing silicon PC waveguide to photonic platforms of other materials, such as silicon nitride, DFG pumped by lasers at the visible and telecom-band range could be examined by the near-infrared photodetectors.

The above results of OFCs are carried out with the pumps of narrowband lasers. The broadband operation of SFG indicates the capability of OFCs pumped with a broadband light. Here, a superluminescent light-emitting diode (SLED) with an emission wavelength ranging from 1535 nm to 1575 nm is chosen, whose emission spectrum is shown in the top panel of Figure 4d. Different from the narrowband laser, the SLED is an incoherent light source and its power spectral intensity is very low. The bottom panel of Figure 4d displays the collected OFC signal of SLED from the top surface of the GaSe-covered PC waveguide region. A broad signal spans from 770 nm to 785 nm is achieved. The halved spectrum of the SLED emission is also overlaid in the dashed line. Though there are several resonant modes between 1535 nm and 1575 nm, the OFC signal of SLED has a moderately uniform conversion efficiency by referring to the spectrum of the SLED, presenting a broadband doubled frequency conversion of SLED. It is governed by the same mechanism as the broadband SFG shown in Figure 4c. With the enhancements from the several resonant modes, the light components at off-resonance wavelengths are involved into the OFCs as well. These results illustrate further the proposed GaSe-induced microcavity on PC waveguide could function as an attractive on-chip optical frequency converter with merits of broad bandwidth and high efficiency.

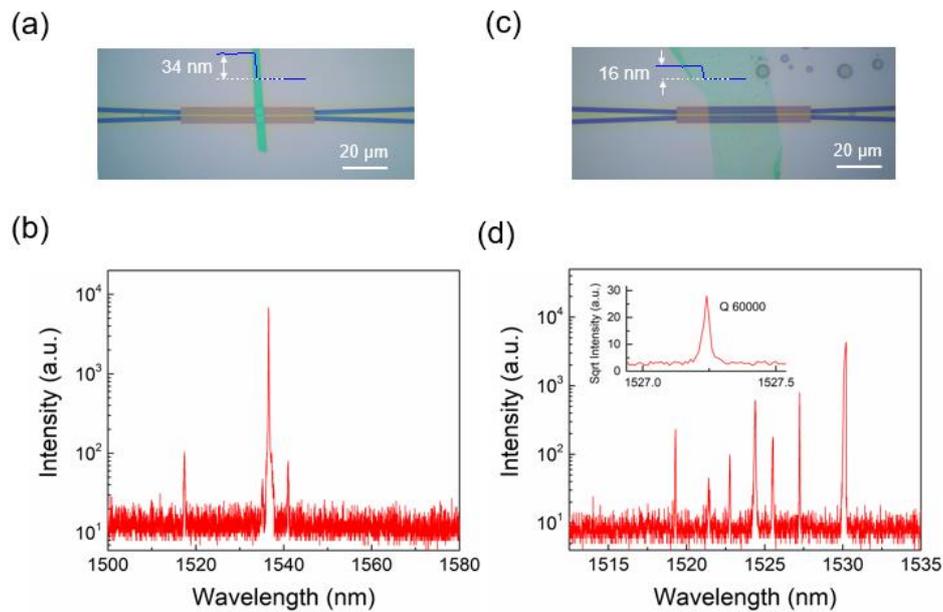

Fig. 5 | Microcavities induced on silicon PC waveguides by few-layer GaSe crystals with different geometries. (a) Optical microscope image of the device covered with a narrow GaSe strip. (b) Wavelength dependent SHG enhanced by the GaSe-induced microcavities in (a). (c) Optical microscope image of the device covered with a thin and wide GaSe film. (d) Wavelength dependent SHG enhanced by the GaSe-induced microcavities in (c). Inset: Wavelength dependence of SHG with square root of intensity around the resonant wavelength of 1527.3 nm, indicating a high $Q$ factor about $6 \times 10^4$.

The characteristics of the microcavity induced by few-layer GaSe crystal on silicon PC waveguide could be tuned by changing the geometries of the integrated GaSe crystals for specific nonlinear optical applications. In Figure 5a and 5c, we show another two devices covered by GaSe flakes with thicknesses of 34 nm and 16 nm and widths of 3.8 μm and 30.6 μm, respectively. The bottom silicon PC waveguide of the two devices is designed and fabricated with the same parameters. To examine the resonant modes of the induced microcavities and the enhanced OFCs, we monitor the scattering SHG signals from the devices with the wavelength of the input narrowband CW laser tuned over a wide range, as displayed in Figure 5b and 5d. From the device in Figure 5a with shorter GaSe strip, only four resonant modes are observed at 1517.38, 1535.18, 1536.58, and 1541.06 nm. It can be explained that when the length of the microcavity induced on the PC waveguide (FP type resonator) is short, the orders of the supported standing modes locating in the PC photonic bandgap is reduced. The maximum $Q$ factor extracted from the presented resonant modes is 12000 at 1536.58 nm, which is smaller than that obtained from the device in Figure 2. The degradation of $Q$ factor is mainly contributed by the strong scattering from the interface between GaSe and the PC waveguide in the short microcavity. By reducing the thickness of GaSe, the mode scattering can be significantly reduced. In the device covered with the 16 nm thick GaSe, as shown in Figure 5d, a maximum $Q$ factor of $6 \times 10^4$ is achieved at the resonant wavelength of 1527.23 nm. The number of resonant modes is more than that in the device shown in Figure 5a because of the enlarged cavity length.

## Conclusions

In conclusion, we have demonstrated the post-induction of microcavity on silicon PC waveguide by integrating a few-layer GaSe crystal, which supports high-efficiency OFCs for providing on-chip light sources with the assistance of GaSe's high second-order nonlinearity. The induced microcavities are formed by lowering the dispersion bands of the waveguide modes with the covered GaSe flake, leading to a mode-gap difference. Benefiting from the small contrast of the mode refractive indices between the bare PC waveguide and GaSe-PC waveguide, the resonant modes could be confined well in the GaSe-covered region by the side bare PC waveguides, which represent high $Q$ factors exceeding $10^4$ reliably. Since the microcavity is composed by the top GaSe flake and the bottom silicon PC waveguide, GaSe is coupled with the resonant mode effectively. Combining with GaSe's high second-order nonlinearity, high-efficiency SHG and SFG are achieved in the integrated device pumped with CW lasers. A record-high conversion efficiency of on-chip SHG up to 131100% $W^{-1}$ is obtained, which promises the directly imaging of the SHG signal by a camera with low sensitivity. In addition, driven by an on-resonance laser, high-efficiency SFG could be realized with off-resonance CW lasers spanning over a broad wavelength range from 1500 to 1580 nm. Broadband OFCs of an incoherent SLED with low spectral power density is also enabled by the cavity-enhancement. The modifications of the induced microcavities by changing the geometries of the integrated GaSe is also demonstrated with improved $Q$ factors or reduced mode numbers.

Our results illustrate that the induced high $Q$ microcavity and high-efficiency OFCs in the GaSe-integrated silicon PC waveguide could provide new strategies to manipulating light-matter interactions, nonlinear photonics, and realizing light sources at varied wavelength ranges in PICs. Other noncentrosymmetric layered materials with high second-order nonlinearity, such as InSe, $NbOCl_2$,[38] could also be utilized to integrated on silicon PC waveguide for the efficient OFCs. With well developments of large-scale growth and transfer of the few-layer crystals,[39,40] it is possible to realize scalable production of the induced microcavities on the silicon photonic platform by selectively post-etching the few-layer crystal into the desired strips.

## Methods

**Device fabrication:** The photonic crystal (PC) waveguide was fabricated with standard electron beam lithography and inductively coupled plasmon etching. The chip was cleaned with piranha solution followed by a DI water rinse to minimize the contamination. Before the transfer of the GaSe layer, the PC waveguide was undercut by buffered oxide etch (BOE) solution. The GaSe layer was exfoliated from their bulk crystals on PDMS in ambient conditions and transferred onto silicon PC waveguide utilizing the dry transfer method in the glove box. A piezo-driven motor was used to control the release speed of samples during the transfer processes as slow as possible. Atomic force microscopy (Bruker Dimension icon) characterizations were performed to measure the thicknesses of the transferred GaSe flakes.

**Optical characterization:** The optical test was performed with a vertical fiber coupling system. For the transmission test, a wavelength-tunable narrowband CW laser (Yenista, T100S-HP/CL) was coupled into one of the grating couplers and monitoring the output power from another grating coupler. A fiber polarization controller was used to maximize the power coupled into the waveguide. For the SHG and SFG test, the monitoring fiber is moved away from the output grating coupler onto the GaSe covered region for collecting the scattering SHG and SFG signals. The spectra of the scattering frequency conversion signals are analyzed by spectrometer. In the SFG test, another wavelength-tunable narrowband CW laser (santec, TSL-710) was used. The SHG image was token by a common CMOS camera equipped in a long working distance monocular zoom stereo microscope, which has a similar sensitivity of the camera equipped in an iPhone. Even though the stereo microscope is designed for aligning the optical fiber with the grating couplers, which has a low numerical aperture, the SHG can be imaged easily proving its high conversion efficiency. To calibrate the absolute emission power of the SHG signal, a commercial 780 nm laser diode with similar light-emitting area as that of the GaSe coverage on the PC waveguide was placed under the same camera for the SHG imaging. By keeping the same imaging conditions, the brightness of the diode was tuned to make sure the integral of optical powers from all pixels of the camera is the same as that from the captured SHG image. Then, the power of the laser diode was measured with a powermeter to determine the SHG power.

## Acknowledgements

This project was primarily supported by the Key Research and Development Program (2022YFA1404800), National Natural Science Foundation of China (12374359, 62104198, 62375225, 62305270), National Key Laboratory Foundation (2023-JCJQ-LB-007), Shaanxi Fundamental Science Research Project for Mathematics and Physics (22JSY004), Xi'an Science and Technology Plan Project (2023JH-ZCGJ-0023). We thank the Analytical & Testing center of Northwestern Polytechnical University for equipment support.

## Author contributions

X.G. conceived and supervised the project. X.C. and Y.Y.Z. carried out the device fabrication and characterization. Y.J. performed the theoretical calculation. Y.Z., J.W., X.H.W. and C.Z. assisted the device fabrication. L.F., B.J. and J.Z. provided valuable discussions and improvements of manuscript. X.C. and Y.Y. Z. contributed equally to the work.

## Competing interests

The authors declare no competing financial interests.

## Supplementary information

Additional information and details on fabrication, characterization of devices and device performance.

# Corresponding Author:

Xuetao Gan is currently a professor working in the School of Physical Science and Technology, Northwestern Polytechnical University. He received his Ph.D. degree (2013) from Northwestern Polytechnical University. Then, he joined the faculty of Northwestern Polytechnical University. His current research interests include 2D materials, nanophotonics and integrated photonics.

Email: xuetaogan@nwpu.edu.cn